\RequirePackage{etex}
\documentclass{iau}
\usepackage{xurl}   

\usepackage{amsmath}
\usepackage{multirow}

\usepackage{epstopdf}
\usepackage{graphicx}
\PassOptionsToPackage{pdftex}{hyperref}
\usepackage{makecell}

\newcommand\aap{A\&A}                
\newcommand\aj{AJ}                   
\newcommand\apj{ApJ}                 
\newcommand\mnras{MNRAS}             

\begin{document}

\lefttitle{Extended Main Sequence Turnoff of NGC\.3532 }
\righttitle{Rao \& Chen}

\jnlPage{1}{7}
\jnlDoiYr{2025}
\doival{10.1017/xxxxx}

\aopheadtitle{Proceedings IAU Symposium}
\editors{C. Sterken,  J. Hearnshaw \&  D. Valls-Gabaud, eds.}

\title{Investigating Unusually Extended Main Sequence Turnoff of the Galactic Open Cluster NGC\,3532}

\author{Khushboo K Rao \thanks{email: khushboo@astro.ncu.edu.tw}, and Wen-Ping Chen}
\affiliation{Institute of Astronomy, National Central University, Taiwan}



\begin{abstract}
Extended main-sequence turnoffs are observed in the color-magnitude diagrams of some young and intermediate-age star clusters. Here we report the case study of NGC\,3532.  Among the cluster member candidates identified using the ML-MOC algorithm on \textit{Gaia} DR3 sources, we found that fast rotators are significantly redder.  Stars near the main-sequence turnoff exhibit a multi-modal distribution in the \textit{Gaia} broadening velocity ($v_{\rm broad}$).  A comparison with relevant literature suggests binaries being typically interspersed with fast rotators, often manifesting as slow to moderate rotators. Additionally, chemically peculiar stars are predominantly located in the bluer regions of the turnoff, with those in the fast-rotator region having large values of \textit{Gaia} re-normalised unit weight error, indicating plausible companions. Our analysis suggests multiple channels responsible for angular momentum loss to account for the extended turnoff.
\end{abstract}


\maketitle

\section{Introduction}

Some young and intermediate-age clusters with ages $< 2$~Gyr exhibit a broader main-sequence turnoff (MSTO) than other evolutionary sequences \citep{Milone2009,Cordoni2018}, a phenomenon termed extended main sequence turnoff (eMSTO). It was first identified in the Magellanic Cloud cluster \citep{Mackey2007}. Related phenomena include split main sequences (MS), multiple MS, and extended MS.

Rotation has emerged as the prevailing explanation for the eMSTO \citep{Bastian2009}, in the sense that stars near the MSTOs in young and intermediate-age clusters have radiative envelopes, thereby hindering efficient spin-down during their MS lifetime and remaining as fast rotators. Recent observations, however, show that these MSTOs comprise both fast and slow rotators \citep{Marino2018b}.  What is the mechanism that allows certain stars to maintain rapid rotation, yet others to slow down? The presence of slow rotators has been attributed to tidal synchronization of binaries, star-disk interactions during pre-MS phases, and stellar mergers.

We have investigated the eMSTO and related phenomena in 53 galactic OCs (Rao et al. in prep). Here, we showcase NGC\,3532, 
a $\sim330$~Myr old open cluster with a heliocentric distance of 490~pc.  With a low extinction ($A_{\text{v}} = 0.07$), its members selected by \textit{Gaia} data exhibit a clear eMSTO that should be little affected by differential reddening in the foreground and, therefore, is intrinsic to the cluster itself.  We present the data and membership selection, followed by analysis and discussion of the results that lead to a tantalizing conclusion. 

\section{Data and Membership} \label{sec:data}

\begin{figure}
\begin{center}
 \includegraphics[width=0.90\textwidth]{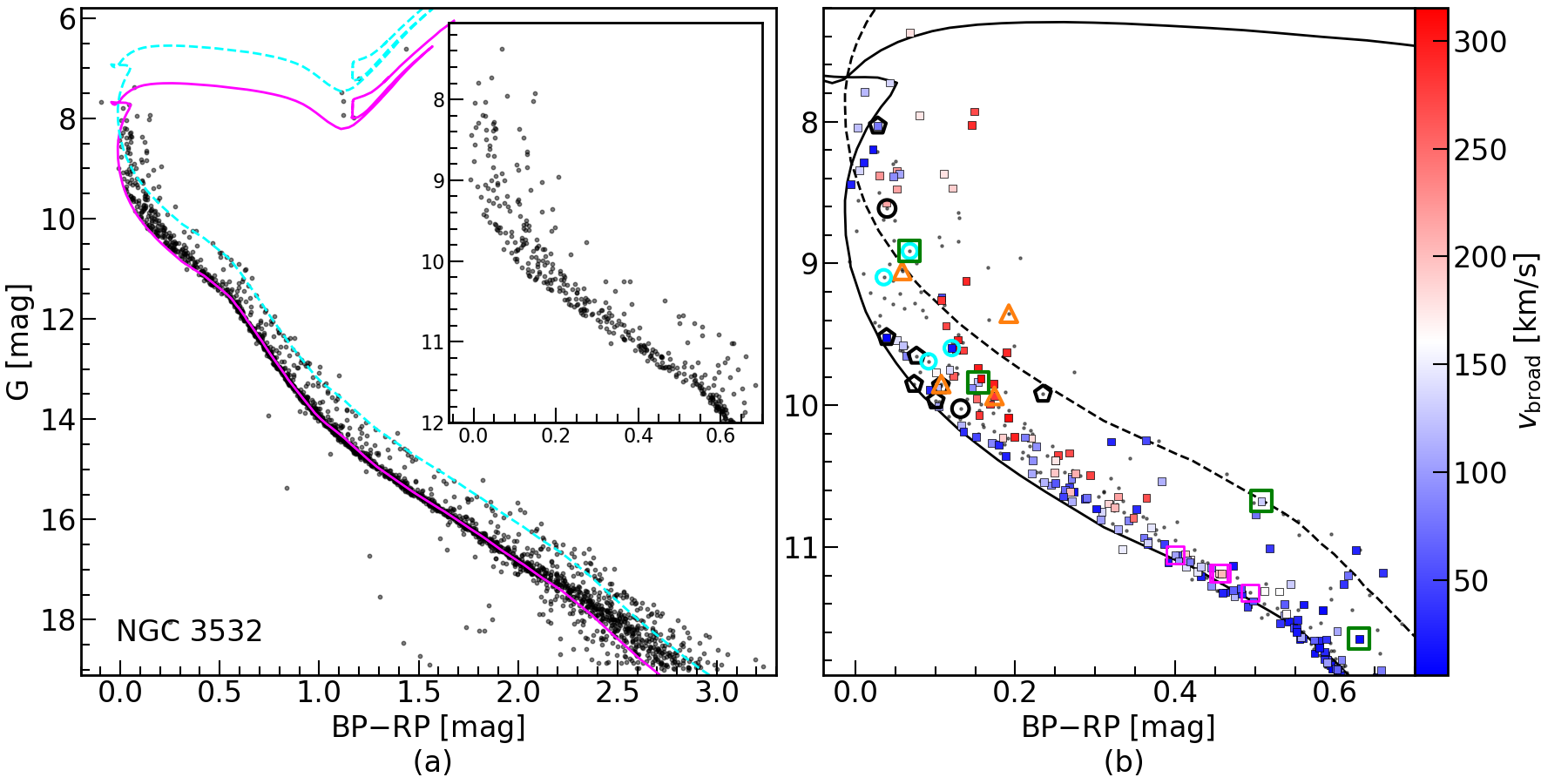} 
 \caption{(a): \textit{Gaia} CMD of NGC\,3532, fitted with a non-rotating isochrone (magenta solid line) of $\log{\rm (Age/year)}= 8.52$, distance of 490~pc, $A_{\text{v}} = 0.07$, and $[Fe/H] = 0.0$. The cyan dashed line shows an equal-mass binary track shifted by $G -0.75$~mag. The inset zooms in to the MSTO region. (b):~CMD of the MSTO stars, color-coded by the $v_{\text{broad}}$ measurements. Grey dots show all the cluster members. The black open circles show eclipsing binaries \citep{Gaia_var2023}. ACV$\mid$CP$\mid$MCP$\mid$ROAM$\mid$ROAP$\mid$SXARI variables \citep{Gaia_var2023} having RUWE$< 1.21$ are shown as black pentagons, and those having RUWE$> 1.21$ as orange triangles. Also marked are eclipsing binaries and spectroscopic binaries \citep[cyan open circles,][]{Gavras_kv_2023}, X-ray sources \citep[green open squares,][]{Chen2023}, and $H_{\alpha}$ sources \citep[magenta open squares,][]{He2025}.  }
   \label{fig1}
\end{center}
\end{figure}

We used the ML-MOC algorithm \citep{Agarwal2021} on \textit{Gaia} DR3 data \citep{GaiaDR32023} to identify cluster members. ML-MOC is a machine learning approach that integrates the k-Nearest Neighbor and Gaussian Mixture Model algorithms, leveraging proper motions and parallaxes to determine cluster membership. We identified a total of 2,266 members within the 200 arcmin field toward the cluster. A total of 2,172 members are within the cluster radius of 120 arcmin. The cluster radius is determined as the distance from the cluster center (RA = 166.396804; Dec: $-58.712294$) at which the member distribution appears to merge with that of the field stars. The CMD of the cluster, shown in Fig.~\ref{fig1}, is fitted with a non-rotating PARSEC isochrone of $\log{\rm (Age)} = 8.52$, distance of 490~pc, metallicity of [Fe/H] = 0.0 dex, and extinction of $Av = 0.07$~mag (Rao et al. in prep). 

\section{Results \& Discussion} \label{sec:results}

In Fig.~\ref{fig1}(a), we see a maximum fraction of fast and slow rotators being located in the redder and bluer regions of the CMD, respectively, indicating the stellar rotation as the significant driver of the eMSTO phenomenon. However, the presence of a small fraction of slow rotators among fast rotators may be indicative of some underlying processes like tidally syncronized binaries. 

\begin{figure}
\begin{center}
 \includegraphics[width=0.4\linewidth]{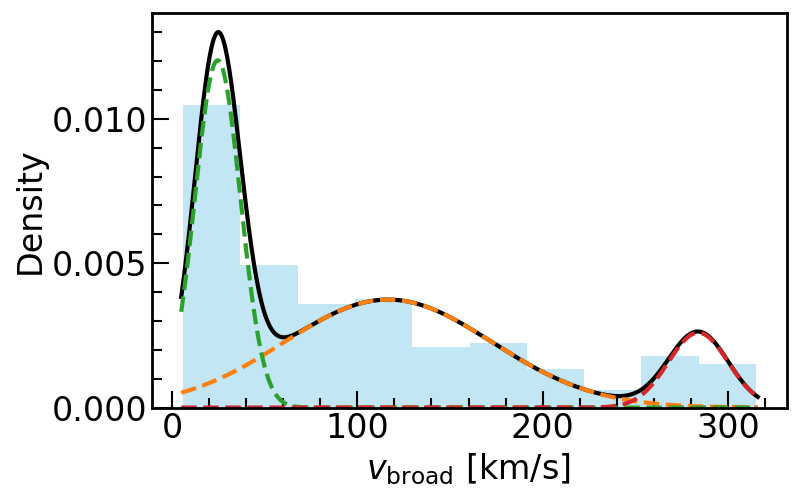}
  \includegraphics[width=0.4\linewidth]{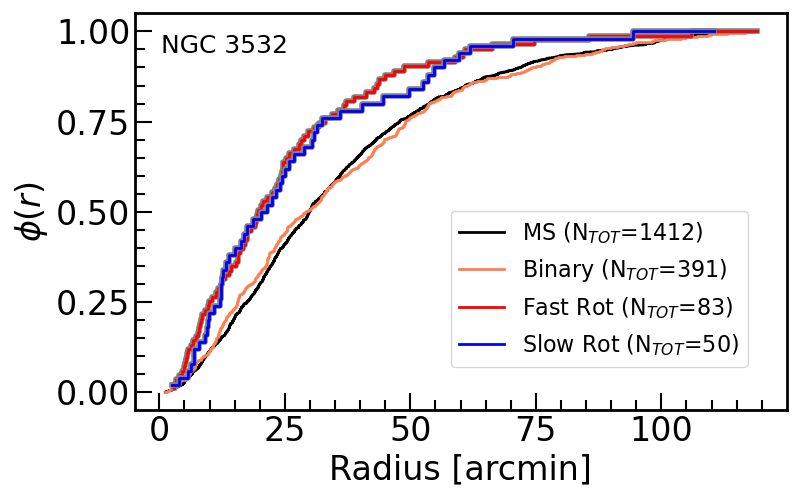} 
 \caption{(Left) $v_{\text{broad}}$  distribution of MSTOs of the cluster having G $< 11.8$~mag. The black solid line represents the fitted multi-modal distribution. The red, green, and orange solid lines represent the fitted Gaussians for individual subsets of the distribution. (Right) Cumulative radial distributions of fast-rotating MSTOs, i.e., $v_{\text{broad}} > 100$~km/s (red solid line), slow-rotating MSTOs having $v_{\text{broad}} < 100$~km/s (blue solid line), high mass-ratio binaries with q $> 0.6$ within primary masses of 0.31 -- 1.5 M$_{\odot}$ (orange solid line), and MS within masses of 0.31 -- 1.5 M$_{\odot}$ (black solid line).}
   \label{fig2}
\end{center}
\end{figure}

The distribution of  $v_{\text{broad}}$ measurements for MSTO stars (MSTOs) with $G < 11.8$~mag, shown in Fig.~\ref{fig2}, reveals a multi-modal distribution pattern composed of three Gaussians. The optimal number of Gaussian components that fit the $v_{\text{broad}}$ distribution best is determined using the Bayesian Information Criterion \citep{1978AnSta...6..461S}. This statistical approach allows for a robust comparison of model Gaussians. This multi-modal distribution suggests that MSTOs of the cluster are experiencing varying degrees of angular momentum retention or loss.

The fast ($v_{\text{broad}} \, > 100$~km/s) and slow rotators ($v_{\text{broad}} \, \leq 100$~km/s) are observed to have similar cumulative radial distributions, as seen in Fig.~\ref{fig2}. Both populations are segregated compared to high-mass ratio binaries ($q > 0.6$) and MS within the mass range of 0.35 M$_{\odot}$ to 1.5 M$_{\odot}$.   

We compared MSTOs of G $< 12$~mag (masses $> 1.3$~M$_{\odot}$) with the literature, and the details are listed in Table~\ref{tabel1}. Four low-mass MSTOs (1.5--1.8~M$_{\odot}$) are found to exhibit H${\alpha}$ emission \citep{He2025} and have $v_{\text{broad}}$ measurements ranging from $102.74\pm11.1$ to $202.8\pm44.52$~km/s, with one of which classified as a $\delta \, \text{Scuti}\mid \gamma \, \text{Doradus}\mid$SX Phoenicis type variable \citep{Gaia_var2023}. H${\alpha}$ emission in these MSTOs may be associated with a decreting disk, a weak magnetic field leading to stellar activity, resulting in angular momentum loss in some cases.  

Furthermore, four MSTOs are found to exhibit X-ray emission \citep{Chen2023}. Of these, two occupy the equal-mass binaries region, with one of them being an eclipsing binary \citep{Gavras_kv_2023}. One X-ray source is located at the low-mass end of the MSTOs, well separated from the MS, and is thus likely to be a high mass-ratio binary, while the other is located in the fast rotators region. Of the four X-ray MSTOs, three are characterized as slow to moderate rotators with v$_{\text{broad}}$) ranging from $13.32\pm13.76$~km/s to $129\pm16.63$~km/s.  The combination of slow rotation, X-ray emission, and their positions on the CMD suggests that these objects are likely binaries, and interactions during the MS or pre-MS phase may have modified their initial angular momentum.

Based on \textit{Gaia} variability catalog \citep{Gaia_var2023}, two sources are classified as short-period (2.19 and 8.19 days) eclipsing binaries and are located in redder regions of the CMDs. A variability catalog, \citet{Gavras_kv_2023}, classified one MSTOs as an eclipsing binary and three MSTOs as short-period (6.298--11.5 days) spectroscopic binaries with null eccentricities. The eclipsing binary lies on the equal-mass binary isochrone track and is also an X-ray source. One of the spectroscopic binaries is a slow rotator with $v_{\text{broad}} = 17.89\pm8.03$~km/s, which is consistent with its being a short-period binary and having a circularized orbit, indicating that it has likely undergone tidal synchronization at some point in its lifetime. Notably, all six binaries are located within the region of fast rotators.

Moreover, the \textit{Gaia} variability catalog \citep{Gaia_var2023} classified 11 MSTOs (2--2.6 M$_{\odot}$) as ACV$\mid$CP$\mid$MCP$\mid$ROAM$\mid$ROAP$\mid$SXARI-type variables, associated with chemically peculiar stars, which are magnetically active and are generally slow to moderate rotators. Of these, seven are located within the slow rotators on the blue edge of the CMD, including one very slow rotator with $v_{\text{broad}} = 9.36\pm5.81$~km/s. This indicates a potential correlation between their variability and rotation. Three of these are located near or redder than the equal-mass binary isochrone; two of these have renormalized unit weight error (RUWE) values of 3.3 and 6.6. One of the 11 is located among the fast rotators, having $v_{\text{broad}} = 280.72\pm113.37$~km/s and RUWE=1.25. A higher RUWE parameter is typically associated with the presence of gravitating companions. This indicates that the chemically peculiar stars residing in the redder regions of CMDs may be part of multiple stellar systems.

\section{Summary \& Conclusion}\label{sec:summary}

In this study, we investigated the Galactic open cluster NGC\,3532 in the context of the eMSTO phenomenon and the presence of slow rotators. We identified six binaries located among the fast-rotating population. Their orbital properties are consistent with being tidally synchronized, and the published $v_{\text{broad}}$ values confirm two of them being indeed slow rotators. We also find that chemically peculiar stars are preferentially located in the bluer region of the CMD, which is dominated by slow rotators. A subset of chemically peculiar stars appears in the redder regions of the CMD but shows elevated RUWE values, suggesting that they are likely members of multiple stellar systems. Our analysis indicates that several channels contribute to angular momentum loss near the MSTO in NGC\,3532, including chemical peculiarity and the evolution of multiple stellar systems during the MS or pre-MS phases.
\begin{table}[!ht]
    \caption{The details of the MSTOs compared with literature.}
    \centering
    {\scriptsize
    \begin{tabular}{ccccccccc}
        Gaia DR3 source id & RA & DEC & RUWE& G & $v_{\text{broad}} \, \pm$ err & Comment & Ref \\
         & (deg) & (deg) & & (mag) & (km/s) &  &    \\ \hline

        5338661808024777856 & 166.386943 & $-$58.730489 & 1.32 & 8.913 & --  & ECL$^a$, X-ray$^b$ & 1, 2 \\ 
        5338654351960981504 & 166.125911 & $-$58.863005 & 0.84 & 9.837 & 119.28$\pm$7.73 & X-ray$^b$ & 2 \\ 
        5338702386868676736 & 165.981678 & $-$58.798258 & 1.16 & 10.676 & 129.26$\pm$16.63 & BIN$^f$, X-ray$^a$ & 2 \\ 
        5338661945464004480 & 166.343199 & $-$58.743262 & 1.05 & 11.646 & 13.32$\pm$13.76  & X-ray$^a$ & 2 \\ 
        5338645654599400704 & 166.642638 & $-$58.845020 & 0.77 & 11.054 & 102.74$\pm$11.1  & H$_{\alpha}^c$ & 3 \\ 
        5338661636226329216 & 166.309856 & $-$58.742543 & 0.74 & 11.184 & 188.72$\pm$47.86  & H$_{\alpha}^c$ & 3 \\ 
        5338624390266300416 & 166.467585 & $-$59.116059 & 0.70 & 11.185 & 202.8$\pm$44.52  & H$_{\alpha}^c$ & 3 \\ 
        5338652118577887616 & 166.365257 & $-$58.861024 & 0.89 & 11.321 & --  & H$_{\alpha}^c$ & 3 \\ 
        5340149241088915712 & 166.713357 & $-$58.706767 & 0.92 & 8.612 & --  & \makecell[c]{ECL$^a$ \\ (q = 0.692, $P$ = 2.1936 d, $e$ = 0)} & 4 (5) \\ 
        5338714962525372160 & 165.860985 & $-$58.497346 & 0.97 & 10.024 & -- &  \makecell[c]{ECL$^a$ \\ (A6V$+$K3V, $P$ = 8.1919 d, $e =$ 0)} & 4 (5) \\ 
        5337865967747145344 & 167.923604 & $-$59.391366 & 0.99 & 8.030 & 79.14$\pm$4.98 &  ACV$^d$ & 4 \\ 
        5337807448781217024 & 168.304663 & $-$59.641640 & 3.89 & 9.053 & -- & ACV$^d$ & 4 \\ 
        5338913596166873344 & 165.080872 & $-$58.407417 & 6.67 & 9.354 & -- & ACV$^d$ & 4 \\ 
        5338645555866749952 & 166.655301 & $-$58.860761 & 0.86 & 9.521 & 9.36$\pm$5.81 &  ACV$^d$ & 4 \\ 
        5340218029295260032 & 166.010235 & $-$58.444511 & 0.85 & 9.652 & -- &  ACV$^d$ & 4 \\ 
        5340165355808070016 & 166.418946 & $-$58.597710 & 1.27 & 9.850 & -- &  ACV$^d$ & 4 \\ 
        5338660807243994112 & 166.534783 & $-$58.670716 & 1.21 & 9.851 & -- &  ACV$^d$ & 4 \\ 
        5338655039155627008 & 166.249908 & $-$58.831366 & 0.82 & 9.862 & -- &  ACV$^d$ & 4 \\ 
        5337757184825229056 & 168.327361 & $-$60.178373 & 1.16 & 9.920 & -- &  ACV$^d$ & 4 \\ 
        5338709121376508160 & 165.808584 & $-$58.596546 & 1.25 & 9.934 & 280.72$\pm$113.37  & ACV$^d$ & 4 \\ 
        5340147797979862912 & 166.911099 & $-$58.790074 & 0.89 & 9.967 & -- &  ACV$^d$ & 4 \\ 
        5338625283619443840 & 166.312769 & $-$59.082678 & 0.94 & 9.097 & -- &  SB$^e\, (e =$ 0) & 1 \\ 
        5337851944644789632 & 167.090170 & $-$59.556335 & 0.88 & 9.597 & 17.89$\pm$8.03 &  SB$^e\, (e =$ 0) & 1 \\ 
        5340155906879318400 & 167.151747 & $-$58.533244 & 1.57 & 9.690 & -- &  SB$^e\, (e =$ 0) & 1 \\ \hline
    \end{tabular}
    }
    \label{tabel1}
    \smallskip 
{\scriptsize $^1$\citet{Gavras_kv_2023}, $^2$\citet{Chen2023}, $^3$\citet{He2025}, $^4$\citet{Gaia_var2023}, $^5$\citet{Ozdarcan_2022}

$^a$Eclipsing binary; $^b$X-ray emitting source; $^cH_{\alpha}$ emission line; $^d$ACV$\mid$CP$\mid$MCP$\mid$ROAM$\mid$ROAP$\mid$SXARI type variable; $^e$spectroscopic binary; $^f$ located on equal mass binary isochrone; $q \, = \, M_2/M_1$; $P\, = $ Period; $e \, = $ eccentricity}
\end{table}


\begin{thebibliography}{}

\bibitem[Mackey \& Broby Nielsen(2007)]{Mackey2007} Mackey, A.~D. \& Broby Nielsen, P.\ 2007, \mnras, 379, 1, 151. 

\bibitem[Cordoni et al.(2018)]{Cordoni2018} Cordoni, G., Milone, A.~P., Marino, A.~F., et al.\ 2018, \apj, 869, 2, 139. 

\bibitem[Milone et al.(2009)]{Milone2009} Milone, A.~P., Bedin, L.~R., Piotto, G., et al.\ 2009, \aap, 497, 3, 755. 

\bibitem[Bastian \& de Mink(2009)]{Bastian2009} Bastian, N. \& de Mink, S.~E.\ 2009, \mnras, 398, 1, L11. 

\bibitem[Marino et al.(2018)]{Marino2018b} Marino, A.~F., Przybilla, N., Milone, A.~P., et al.\ 2018, \aj, 156, 3, 116. 

\bibitem[He et al.(2025)]{He2025} He, C., Li, C., \& Li, G.\ 2025, \apj, 979, 2, 246. 

\bibitem[Chen et al.(2023)]{Chen2023} Chen, S., Kargaltsev, O., Yang, H., et al.\ 2023, \apj, 948, 1, 59. 

\bibitem[Eyer et al.(2023)]{Gaia_var2023} Eyer, L., Audard, M., Holl, B., et al.\ 2023, \aap, 674, A13. 

\bibitem[Gavras et al.(2023)]{Gavras_kv_2023} Gavras, P., Rimoldini, L., Nienartowicz, K., et al.\ 2023, \aap, 674, A22. 

\bibitem[{\"O}zdarcan(2022)]{Ozdarcan_2022} {\"O}zdarcan, O.\ 2022, \mnras, 509, 2, 1912. 

\bibitem[Agarwal et al.(2021)]{Agarwal2021} Agarwal, M., Rao, K.~K., Vaidya, K., et al.\ 2021, \mnras, 502, 2, 2582. 

\bibitem[Gaia Collaboration et al.(2023)]{GaiaDR32023} Gaia Collaboration, Vallenari, A., Brown, A.~G.~A., et al.\ 2023, \aap, 674, A1. 

\bibitem[Schwarz(1978)]{1978AnSta...6..461S} Schwarz, G.\ 1978, Annals of Statistics, 6, 2, 461. 

\end{thebibliography}
\end{document}